\documentclass[conference]{IEEEtran}
\IEEEoverridecommandlockouts

\usepackage{cite}
\usepackage{amsmath,amssymb,amsfonts}

\usepackage{bbm}
\usepackage{graphicx}
\usepackage{textcomp}
\usepackage{xcolor}
\usepackage{kotex}
\usepackage{algorithmicx,algpseudocode}

\usepackage{booktabs}
\usepackage[linesnumbered,ruled]{algorithm2e}
\usepackage{enumitem}   
\usepackage{eso-pic}
\usepackage{booktabs}
\usepackage{float}

\def\BibTeX{{\rm B\kern-.05em{\sc i\kern-.025em b}\kern-.08em
    T\kern-.1667em\lower.7ex\hbox{E}\kern-.125emX}}
\begin{document}

\title{Neural Architectural Nonlinear Pre-Processing for mmWave Radar-based Human Gesture Perception}

\author{
\IEEEauthorblockN{Hankyul Baek}
\IEEEauthorblockA{\textit{Korea University}
\\Seoul, Korea
\\ 67back@korea.ac.kr}
\and
\IEEEauthorblockN{Yoo Jeong (Anna) Ha}
\IEEEauthorblockA{\textit{Korea University}
\\Seoul, Korea
\\ annaha17@korea.ac.kr}
\and
\IEEEauthorblockN{Minjae Yoo}
\IEEEauthorblockA{\textit{Korea University}
\\Seoul, Korea
\\ mj7015@korea.ac.kr}
\and
\IEEEauthorblockN{Soyi Jung}
\IEEEauthorblockA{\textit{Ajou University}
\\Suwon, Korea
\\ sjung@ajou.ac.kr}
\and
\IEEEauthorblockN{Joongheon Kim}
\IEEEauthorblockA{\textit{Korea University}
\\Seoul, Korea
\\ joongheon@korea.ac.kr}
}

\maketitle

\begin{abstract}
In modern on-driving computing environments, many sensors are used for context-aware applications. This paper utilizes two deep learning models, U-Net and EfficientNet, which consist of a convolutional neural network (CNN), to detect hand gestures and remove noise in the Range Doppler Map image that was measured through a millimeter-wave (mmWave) radar. To improve the performance of classification, accurate pre-processing algorithms are essential. Therefore, a novel pre-processing approach to denoise images before entering the first deep learning model stage increases the accuracy of classification. Thus, this paper proposes a deep neural network based high-performance nonlinear pre-processing method.
\end{abstract}


\section{Introduction}\label{sec1}
Endless research is conducted by countless institutes to bring self-driving vehicles to the mainstream market~\cite{pieee202105park,ijcai2019shin}. Simulations to evaluate the performance of this improved denoising algorithm are implement on human gesture datasets. It is emphasized that this particular algorithm can be directly transferred to the on-driving setting where denoised-signals from the drivers is pivotal to detect for safety measures~\cite{DBLP:conf/icoin/YuL22}. 

\begin{figure}[t!]
    \centering
    \includegraphics[width =0.8\columnwidth]{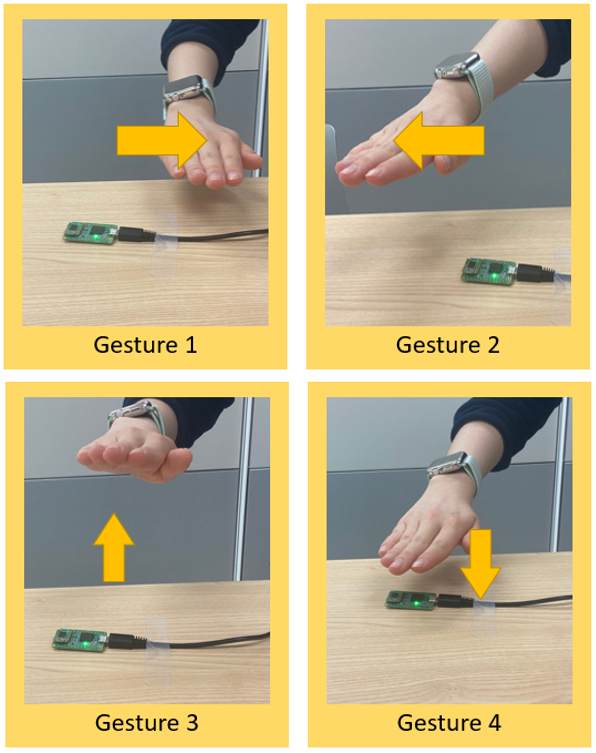}
    \caption{Four way hand gesture recognition data set configuration method using mmWave radar. (From the left, Gesture 1: left, Gesture 2: right, Gesture 3: away from radar, Gesture 4: towards radar).}
    \label{fig:1}
\end{figure}

Among various approaches in human gesture recognition, this paper considers millimeter-wave (mmWave) radar based signal detection and estimation~\cite{jcn2014joongheon,6884155,tvt202108jung,tbc2013joongheon,tvt2016joongheon,iotj17kim,DBLP:conf/icoin/Kim22,9194445,9042896,7790886,7109892,7079536}, as shown in Fig.~\ref{fig:1}.
The reasons why mmWave radar-based approach is considered are because it works well even in dark situations where visual sensing modules cannot work properly~\cite{isl19zhang,isj21almalioglu,access20zhao}.

mmWave radar is a type of radar system that transmits electromagnetic wave signals in the millimeter range that provides information on range, velocity and angle of the detected object. mmWave radars are probable sensors with its broad portfolio that allow for minor movements from the reflected surface to be detected. The radar used in this experiment has a lower bound of $57.5$\,GHz and an upper bound of $63.5$\,GHz, giving it a $6$\,GHz signal bandwidth for an even precise detection of movement~\cite{cm08singh,cm14carlos}. mmWave radars can be used alongside convolutional neural network (CNN) to detect human motions in various settings~\cite{ref1,DBLP:conf/icoin/Na21}. 
Along with its precise accuracy, mmWave radars are low-cost, low in energy consumption and compact, making it an ever more promising sensor in autonomous vehicles and highlights the necessity in enhancing its performance.

An original approach to denoise hand gesture motion by synthesising noise images with ground truth images is explained in this paper. This pre-processed image is fed as the input to the combined deep learning U-Net and EfficientNet models. 
By using these deep neural architectures, the benefit of nonlinear pre-processing can be realized.
Through this unique pre-processing method, a higher accuracy in classifying human gestures can be achieved. Possessing a sound pre-processing image obtained from mmWave radars is essential when applying this sensor in diverse scenarios. 

\section{Neural Architectural Nonlinear Pre-Processing for Human Gesture Recognition}\label{sec:proposed}

\subsection{Observation}
To perform the object recognition process, the data obtained from the mmWave radar is configured as a Range Doppler Map. When recognizing motion using mmWave radar, there are a couple of major considerations: 
\begin{itemize}
\item Even with the same signal, the measured noise intensity varies depending on the manufactured sensor and its three antennas.
\item The measurement threshold needs to be adjusted depending on the environment and purpose. \end{itemize}

 \begin{figure}[t]
    \centering
    \begin{tabular}{@{}cc@{}}
         \includegraphics[width=.48\columnwidth]{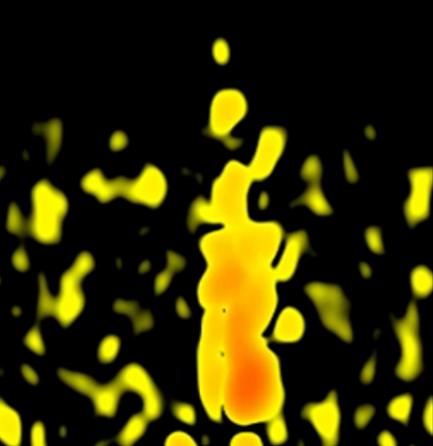}&
         \includegraphics[width=.48\columnwidth]{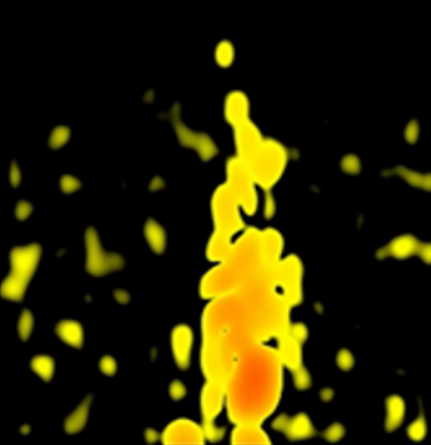}\\
         \small (a) Antenna A. & \small (b) Antenna B.\\
    \end{tabular}
    
     \caption{From the left of the Range Doppler Map, antenna A and antenna B show completely different signals even from the same hand motion.}       
         \label{fig:2}
\end{figure}

 \begin{figure}[t]
    \centering
    \begin{tabular}{@{}cc@{}}
         \includegraphics[width=.48\columnwidth]{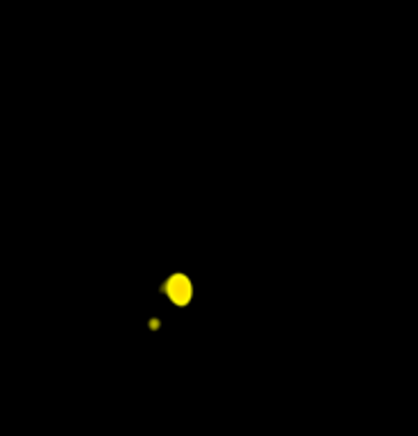}&
         \includegraphics[width=.48\columnwidth]{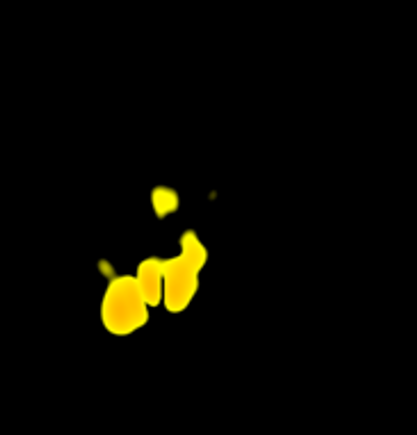}\\
         \small (a) Threshold: -70\,dB. & \small (b) Threshold: -85\,dB.\\
    \end{tabular}
    
     \caption{Range Doppler Map of the same hand motion signal measured through different thresholds.}       
         \label{fig:3}
\end{figure}
\begin{figure}[t]
    \centering
    \includegraphics[width =1\columnwidth]{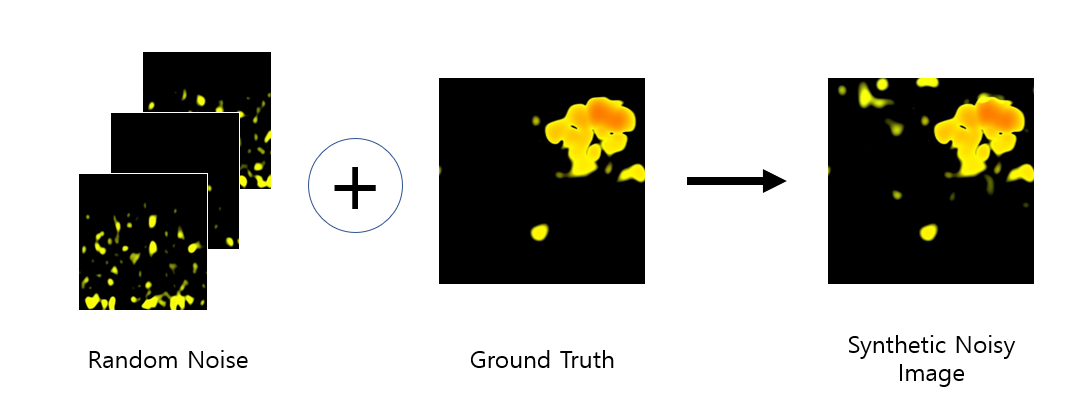}
    \caption{Synthetic Noise Generation Method. Random Noise generated in two thresholds 
    $-95$\,dB and $-100$\,dB is augmented with pure hand motion image to produce Synthetic Noisy Image.}
    \label{fig:sng}
\end{figure}

\begin{figure*}[t]
    \centering
    \includegraphics[width =2.0\columnwidth]{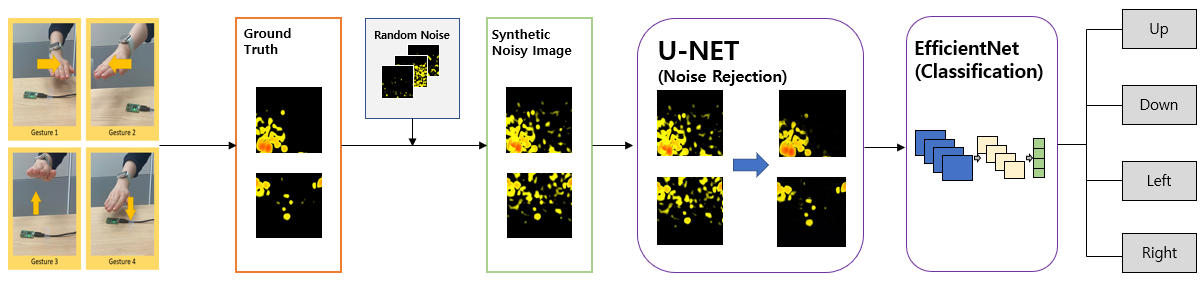}
    \caption{End-to-end neural architectural system model for human gesture recognition.}
    \label{fig:model}
\end{figure*}

As portrayed in Fig.~\ref{fig:2}, it can be observed that even with the same gesture, same signal, the same environment and setting, the noise intensity and its distribution varies profusely depending on received antenna. To counterbalance this phenomenon, deep learning image processing technique, which is robust to variations in noise according to environment and manufactured devices, is adapted.

In addition, as shown in Fig.~\ref{fig:3}, if the threshold value is increased to minimize noise, the original signal may be defected, and all possibilities to decipher the signal may be unavailable. To improve these flaws, an image processing technique using deep learning is adopted to reduce the discrepancies between the antennas of the same device. Hence, by utilizing this pre-processing algorithm, mmWave radar sensors can secure a place in a wide variety of applications where recognition of motion is key.

However, there is one problem that arises when implementing the model introduced in Fig.~\ref{fig:3}. It is impossible to remove noise from the Range Doppler Map image and extract an image to be set as the ground truth simultaneously when using only one antenna. On the other hand, as seen in Fig.~\ref{fig:2}, if more than two antennas are used, discrepancies occur between the Range Doppler Map images even with the same motion applied. Hence, it is impossible to interpret the images and process them further to remove noise from the image.  

\subsection{Proposed Algorithm}
Based on the observation, this paper proposes a novel method to overcome this enigma. The synthetic noise generation technique creates an artificial noise image by combining a ground truth image, which is obtained by sensing hand motion at a $-90$\,dB threshold, and the noise image measured in the absence of motion as shown in Fig.~\ref{fig:sng}, and then training it in the noise removal model. To prevent the operation from learning only a specific threshold noise, one random sample from a mixture batch of approximately $10,000$ noise images measured at two different thresholds, $-95$\,dB and $-100$\,dB values, is combined with the ground truth Range Doppler Map image to create a Synthetic Noisy Image.

The structure of the model proposed in this paper is demonstrated in Fig.~\ref{fig:model}. To emulate the real-life situation of various environments, data containing noise generated when an object is recognized through the mmWave radar is transferred as the input to the pre-trained U-Net. After that, the output from the U-Net model, which would be the denoised Doppler Range image, is fed into the EfficientNet~\cite{enet} to recognize the specific motion of the hand gesture. When the above model is applied to the dataset from the radar, the image with a substantial number of noises is inputted into the U-Net model. This model outputs a denoised image, which then enters the classification model, thereby ensuring the accuracy of motion recognition and eliminating any possible signal damaging due to excessive threshold setting. The proposed model successfully classifies the motion out of the four hand gestures: left, right, moving towards and away from the radar.

 \begin{figure}[t]
    \centering
    \begin{tabular}{@{}cc@{}}
         \includegraphics[width=.48\columnwidth]{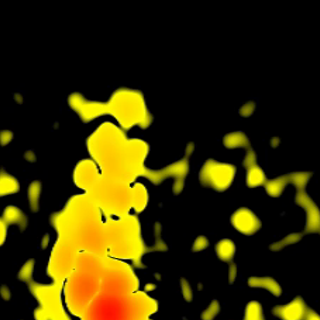}&
         \includegraphics[width=.48\columnwidth]{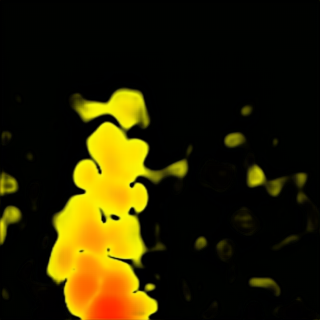}\\
         \small (a)Before. & \small (b) After.\\
    \end{tabular}
    
     \caption{Evaluation for noise rejection.}       
         \label{fig:noisea}
\end{figure}

This model has the characteristic of improving the performance according to its purpose by properly setting the depth, width, and resolution~\cite{enet}.
In this study, EfficientNet-B0, the simplest structure among several EfficientNets, is used.

\section{Experiment}\label{sec:experiment}
The proposed algorithm in this paper is implemented with Infineon's BGT60TR13C 60\,GHz mmWave frequency modulated continuous wave (FMCW) radar. Furthermore, real-world datasets are gathered, generated, and used for the performance evaluation. 

In order to confirm the effectiveness of the motion recognition algorithm mentioned in Fig.~\ref{fig:model}, the experiment is composed as follows. As mentioned, Infineon's BGT60TR13C module is used to implement a mmWave radar motion recognition algorithm that identifies objects after removing noise from the surrounding environments. By setting the threshold values of the mmWave radar to $-95$\,dB and $-100$\,dB, N noise images are extracted for each threshold value. 

As shown in Fig.~\ref{fig:1}, each dataset is created by using the mmWave radar to classify hand movements in four different directions. Subsequently, a Synthetic Noise image is created by combining the Range Doppler Map of the four-direction hand motion signals and the noise image randomly extracted using the Synthetic Noise Generation method. 

The Synthetic Noisy Image generated from the Synthetic Noise Generation is fed into the U-Net. Thereafter, the denoised image outputted from the U-Net is used as the input to the EfficientNet, which then classifies that hand gestures out of the four motions.

Our noise rejection performance is visually presented in Fig.~\ref{fig:noisea}. As can be seen in Fig.~\ref{fig:noisea}, deep learning based approach can remove certain amounts of noises. 
Furthermore, the existing noises in Fig.~\ref{fig:noisea} are real noises from mmWave FMCW radar (not synthetic noises). Therefore, we can obviously confirm that our proposed neural architectural nonlinear pre-processing algorithm works well in real-world noise situations.

\begin{figure}[t!]
    \centering
    \includegraphics[width =0.99\columnwidth]{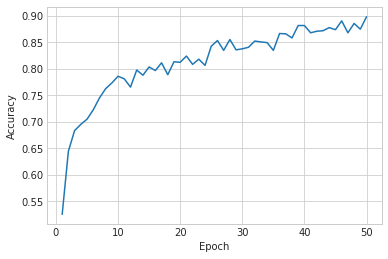}
    \caption{Performance evaluation result.}
    \label{fig:sim}
\end{figure}

Our performance evaluation result in terms of perception accuracy can be obtained as in Fig.~\ref{fig:sim}.
As shown in Fig.~\ref{fig:sim}, we can verify that our proposed neural architectural nonlinear pre-processing converges to the accuracy of (approximately more than) $90$\,\%.

\section{Conclusions}\label{sec:conclusions}
This paper introduced a novel denoising algorithm that allows the mmWave radar to be employed in a diverse environment. The process of concatenating a noise data with a ground truth image to create a Synthetic Noisy Image is the highlight of this paper. This auxiliary step of using a pre-processed, denoised image as the input to a CNN model, produces a much higher accurate level of hand motion classification than compared to that of a conventional method where only raw image with noise was the input to the CNN model. The efficacy of the above object recognition method using deep learning denoising algorithm is confirmed through its ability to successfully classify the four distinct hand movements. The proposed method is promising especially in environments where noise is abundant such as roads for autonomous vehicles. 

\bibliographystyle{IEEEtran}
\bibliography{ref_mmave,ref_aimlab}

\end{document}